\documentclass[twocolumn,floatfix,showpacs,superscriptaddress,eqsecnum]{revtex4}
\usepackage{graphicx}
\usepackage{amsmath}
\usepackage{bm}
\begin{document}

\title{Finite difference method for transport properties of massless Dirac fermions}
\author{J. Tworzyd{\l}o}
\affiliation{Institute of Theoretical Physics, Warsaw University, Ho\.{z}a 69, 00--681 Warsaw, Poland}
\author{C. W. Groth}
\affiliation{Instituut-Lorentz, Universiteit Leiden, P.O. Box 9506, 2300 RA Leiden, The Netherlands}
\author{C. W. J. Beenakker}
\affiliation{Instituut-Lorentz, Universiteit Leiden, P.O. Box 9506, 2300 RA Leiden, The Netherlands}
\date{October 2008}
\begin{abstract}
We adapt a finite difference method of solution of the two-dimensional massless Dirac equation, developed in the context of lattice gauge theory, to the calculation of electrical conduction in a graphene sheet or on the surface of a topological insulator. The discretized Dirac equation retains a single Dirac point (no ``fermion doubling''), avoids intervalley scattering as well as trigonal warping, and preserves the single-valley time reversal symmetry (= symplectic symmetry) at all length scales and energies --- at the expense of a nonlocal finite difference approximation of the differential operator. We demonstrate the symplectic symmetry by calculating the scaling of the conductivity with sample size, obtaining the logarithmic increase due to antilocalization. We also calculate the sample-to-sample conductance fluctuations as well as the shot noise power, and compare with analytical predictions.
\end{abstract}
\maketitle

\section{Introduction}
\label{intro}

The discovery of graphene \cite{Gei07} has created a need for efficient numerical methods to calculate transport properties of massless Dirac fermions. The two-dimensional massless Dirac equation (or Weyl equation) that governs the low-energy and long-wave length dynamics of conduction electrons in graphene has a time reversal symmetry called \textit{symplectic}  --- which is special because it squares to $-1$. (The usual time reversal symmetry, which squares to $+1$, is called \textit{orthogonal}.) The symplectic symmetry is at the origin of some of the unusual transport properties of graphene \cite{Lud94,Ost07,Ryu07,Bee08,Eve08}, including the absence of back scattering \cite{And98}, weak antilocalization \cite{Suz02}, enhanced conductance fluctuations \cite{Kha08,Kec08}, and absence of a metal-insulator transition \cite{Bar07,Nom07}.

Numerical methods of solution can be divided into two classes, depending on whether they break or preserve the symplectic symmetry. 

The tight-binding model of graphene, with nearest neighbor hopping on a honeycomb lattice, breaks the symplectic symmetry by the two mechanisms of intervalley scattering \cite{Suz02} and trigonal warping \cite{McC07}. Intervalley scattering couples the two flavors of Dirac fermions, corresponding to the two different valleys (at opposite corners of the Brillouin zone) in the graphene band structure, thereby changing the symmetry class from symplectic to orthogonal. Trigonal warping is a triangular distortion of the conical band structure that breaks the momentum inversion symmetry ($+\bm{p}\rightarrow-\bm{p}$), thereby effectively breaking time reversal symmetry in a single valley and changing the symmetry class from symplectic to unitary. 

Breaking of the symplectic symmetry eliminates both weak antilocalization as well as the enhancement of the conductance fluctuations, and drives the system to an insulator with increasing size or disorder \cite{Ale06,Alt06}. As observed in computer simulations \cite{Ryc07,Lew08,Zha08}, the breaking of the symplectic symmetry can be pushed to larger system sizes and larger disorder strengths by reducing the lattice constant (at fixed correlation length and fixed amplitude of the disorder potential) --- but this severely limits the computational efficiency. 

The Chalker-Coddington network model \cite{Cha88,Ho96,Kra05}, applied to graphene in Ref.\ \cite{Sny08}, has a single flavor of Dirac fermions, so there is no intervalley scattering --- but it still belongs to the same class of methods that break the symplectic symmetry of the massless Dirac equation. (The symplectic symmetry is broken on short length scales by the Aharonov-Bohm phases that appear in the mapping of the Dirac equation onto the network model.) 

Both the network model and the tight-binding model are \textit{real space} regularizations of the Dirac equation, with a smallest length scale (the lattice constant) to cut off the unbounded spectrum at large positive and large negative energies. There exists at present just one method to calculate transport properties numerically while preserving the symplectic symmetry, developed independently (and implemented differently) in Refs.\ \cite{Bar07} and \cite{Nom07}. That method (used also in Refs.\ \cite{San07,Nom08}) is based on a \textit{momentum space} regularization, with a cutoff of the Fourier transformed Dirac equation at some large value of momentum. 

It is the purpose of the present paper to develop and implement an alternative method of solution of the Dirac equation, that shares with the tight-binding and network models the convenience of a formulation in real space rather than momentum space, but without breaking the symplectic symmetry.

A celebrated no-go theorem \cite{Nie81} in lattice gauge theory forbids any regularization of the Dirac equation with \textit{local} couplings from preserving symplectic symmetry. (The problematic role of intervalley scattering appears in that context as the fermion doubling problem.) Several \textit{nonlocal} finite difference methods have been proposed to work around the no-go theorem and we will adapt one of these (developed by Stacey \cite{Sta82} and by Bender, Milton, and Sharp \cite{Ben83}) to the study of transport properties. 

The adaptation amounts to 1) the inclusion of a spatially dependent electrostatic potential (which breaks the chiral symmetry that played a central role in Refs.\ \cite{Sta82,Ben83}), and 2) a proper discretization of the current operator (such that the total current through any cross section is conserved). We implement the finite difference method to solve the scattering problem of Dirac fermions in a disordered potential landscape connected to ballistic leads, and compare our numerical results for the scaling and statistics of conductance and shot noise power with analytical theories \cite{Kha08,Kec08,Sch08}.

Our numerical method is relevant for electrical conduction in graphene under the assumption that the impurity potential in the carbon monolayer is long-ranged (so that intervalley scattering is suppressed) and weak (so that trigonal warping can be neglected). Massless Dirac fermions are also expected to govern the electrical conduction along the surface of a three-dimensional topological insulator \cite{Fu07,Moo07,Roy07} (recently realized in BiSb \cite{Hsi08}). In that case the symplectic symmetry is preserved even for short-range scatterers, and our numerical results should be applicable more generally.

\section{Finite difference representation of the transfer matrix}
\label{finitedifference}

\subsection{Dirac equation}
\label{Diracequation}

We consider the two-dimensional massless Dirac equation,
\begin{equation}
H\Psi=E\Psi,\;\;H=-i\hbar v(\sigma_{x}\partial_{x}+\sigma_{y}\partial_{y})+U(\bm{r}),\label{Dirac}
\end{equation}
where $v$ and $E$ are the velocity and energy of the Dirac fermions, $U(x,y)$ is the electrostatic potential landscape, and $\Psi(x,y)$ is the two-component (spinor) wave function. The two spinor components of $\Psi$ refer to the two atoms in the unit cell in the application to graphene, or to the two spin degrees of freedom in the application to the surface of a topological insulator. We note the symplectic symmetry of the massless Dirac Hamiltonian,
\begin{equation}
H={\cal S}H{\cal S}^{-1}=\sigma_{y}H^{\ast}\sigma_{y}.\label{symplectic}
\end{equation}
The time reversal symmetry operator ${\cal S}=i\sigma_{y}{\cal C}$ (with ${\cal C}$ the operator of complex conjugation) squares to $-1$. The chiral symmetry $\sigma_{z}H\sigma_{z}=-H$ is broken by a nonzero $U$, so it will play no role in what follows. For later use we also note the current operator
\begin{equation}
(j_{x},j_{y})=v(\sigma_{x},\sigma_{y}).\label{jdef}
\end{equation}

We consider a strip geometry of length $L$ along the longitudinal $x$ direction and width $W$ along the transversal $y$ direction. For the discretization we use a square lattice, $x_{m}=m\Delta$, $y_{n}=n\Delta$, with indices $m=1,2,\ldots M$ ($M=L/\Delta$), $n=1,2,\ldots N$ ($N=W/\Delta$). In the applications we will consider large aspect ratios $W/L\gg 1$, for which the precise choice of boundary conditions in the transverse direction does not matter. We choose periodic boundary conditions, $y_{N+1}\equiv y_{1}$, since they preserve the symplectic symmetry. The values $\Psi_{m,n}=\Psi(x_{m},y_{n})$ of the wave function at a lattice point are collected into a set of $N$-component vectors $\bm{\Psi}_{m}=(\Psi_{m,1},\Psi_{m,2},\ldots\Psi_{m,N})^{T}$, one for each $m=1,2,\ldots M$. 

The $N\times N$ transfer matrix ${\cal M}_{m}$ is defined by
\begin{equation}
\bm{\Psi}_{m+1}={\cal M}_{m}\bm{\Psi}_{m}.\label{transfermatrix}
\end{equation}
The symplectic symmetry \eqref{symplectic} of the Hamiltonian requires that $\Psi$ and $\sigma_{y}\Psi^{\ast}$ are both solutions at the same energy $E$, so they should both satisfy Eq.\ \eqref{transfermatrix}. The corresponding condition on the transfer matrix is
\begin{equation}
{\cal M}_{m}=\sigma_{y}{\cal M}_{m}^{\ast}\sigma_{y}.\label{Msymplectic}
\end{equation}
The transfer matrix should conserve the total current through any cross section of the strip. In terms of the (still to be determined) discretized current operator $J_{x}$, this condition reads $\langle\bm{\Psi}_{m+1}|J_{x}|\bm{\Psi}_{m+1}\rangle=\langle\bm{\Psi}_{m}|J_{x}|\bm{\Psi}_{m}\rangle$, which then corresponds to the following condition on the transfer matrix:
\begin{equation}
{\cal M}_{m}^{\dagger}J_{x}{\cal M}_{m}=J_{x}.\label{currentconservation}
\end{equation}

Our problem is to discretize the differential operators in the Dirac equation \eqref{Dirac}, as well as the current operator \eqref{jdef}, in such a way that the resulting transfer matrix describes a single flavor of Dirac fermions and without violating the two conditions \eqref{Msymplectic}, \eqref{currentconservation} of symplectic symmetry and current conservation.

\subsection{Discretization}
\label{discrete}

A local replacement of the differential operators $\partial_{x},\partial_{y}$ by finite differences either violates the Hermiticity of $H$ (thus violating conservation of current) or breaks the symplectic symmetry (by the mechanism of fermion doubling). A nonlocal finite difference method that preserves the Hermiticity and symplectic symmetry of $H$ was developed by Stacey \cite{Sta82} and by Bender, Milton, and Sharp \cite{Ben83}. These authors considered the case $U=0$, when both symplectic and chiral symmetry are present. We extend their method to a spatially dependent $U$ (thereby breaking the chiral symmetry), and obtain the discretized transfer matrix and current operator.

Since the transfer matrix relates $\Psi(x,y)$ at two different values of $x$, it is convenient to isolate the derivative with respect to $x$ from the Dirac equation \eqref{Dirac}. Multiplication of both sides by $(i/\hbar v)\sigma_{x}$ gives
\begin{equation}
\partial_{x}\Psi=(-i\sigma_{z}\partial_{y}-i\sigma_{x}V)\Psi,\label{Dirac2}
\end{equation}
with the definition $V=(U-E)/\hbar v$. We can now make contact with the discretization in Refs.\ \cite{Sta82,Ben83} of the Dirac equation in one space and one time dimension, with $x$ playing the role of (imaginary) time and $y$ being the spatial dimension.

\begin{figure}[tb]
\centerline{\includegraphics[width=0.8\linewidth]{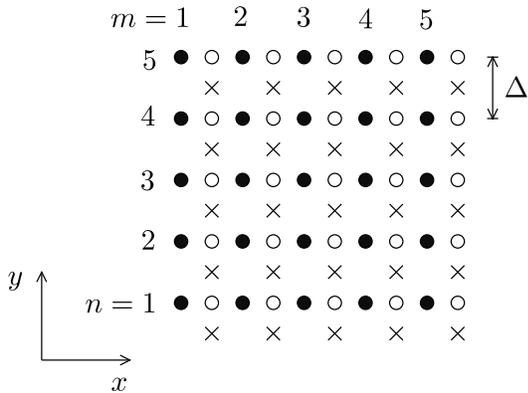}}
\caption{\label{fig_lattice}
Square lattice (filled circles) on which the wave function $\Psi$ is discretized as $\Psi_{m,n}$. The finite differences are evaluated at the displaced points indicated by crosses. The Dirac equation \eqref{Dirac2} is applied at the empty circles, by taking the mean of the contributions from the two adjacent crosses. The resulting finite difference equation defines a transfer matrix in the $x$ direction that conserves current and preserves the symplectic symmetry.}
\end{figure}

The key step by which Refs.\ \cite{Sta82,Ben83} avoid fermion doubling is the evaluation of the finite differences on a lattice that is displaced symmetrically from the original lattice. The displaced lattice points $(x_{m}+\Delta/2,y_{n}+\Delta/2)$ are indicated by crosses in Fig.\ \ref{fig_lattice}. On the displaced lattice, the differential operators are discretized by
\begin{align}
&\partial_{x}\Psi\rightarrow\frac{1}{2\Delta}(\Psi_{m+1,n}+\Psi_{m+1,n+1}-\Psi_{m,n}-\Psi_{m,n+1}),\label{partialx}\\
&\partial_{y}\Psi\rightarrow\frac{1}{2\Delta}(\Psi_{m,n+1}+\Psi_{m+1,n+1}-\Psi_{m,n}-\Psi_{m+1,n}),\label{partialy}
\end{align}
and the potential term is replaced by
\begin{equation}
V\Psi\rightarrow\tfrac{1}{4}V_{m,n}(\Psi_{m+1,n}+\Psi_{m+1,n+1}+\Psi_{m,n}+\Psi_{m,n+1}),\label{VPsi}
\end{equation}
with $V_{m,n}=V(x_{m}+\Delta/2,y_{n}+\Delta/2)$. The Dirac equation \eqref{Dirac2} is applied at the points $(x_{m}+\Delta/2,y_{n})$ (empty circles in Fig.\ \ref{fig_lattice}) by averaging the terms at the two adjacent points $(x_{m}+\Delta/2,y_{n}\pm\Delta/2)$.

The resulting finite difference equation can be written in a compact form with the help of the $N\times N$ tridiagonal matrices ${\cal J}$, ${\cal K}$, ${\cal V}^{(m)}$, defined by the following nonzero elements:
\begin{align}
&{\cal J}_{n,n}=1,\;\;{\cal J}_{n,n+1}={\cal J}_{n,n-1}=\tfrac{1}{2},\label{Idef}\\
&{\cal K}_{n,n+1}=\tfrac{1}{2},\;\;{\cal K}_{n,n-1}=-\tfrac{1}{2},\label{Kdef}\\
&{\cal V}^{(m)}_{n,n}=\tfrac{1}{2}(V_{m,n}+V_{m,n-1}),\;\;{\cal V}^{(m)}_{n,n+1}=\tfrac{1}{2}V_{m,n},\nonumber\\
&{\cal V}^{(m)}_{n,n-1}=\tfrac{1}{2}V_{m,n-1}.\label{Vdef}
\end{align}
In accordance with the periodic boundary conditions, the indices $n\pm 1$ should be evaluated modulo $N$. Notice that ${\cal J}$ and ${\cal V}^{(m)}$ are real symmetric matrices, while ${\cal K}$ is real antisymmetric. Furthermore ${\cal J}$ and ${\cal K}$ commute, but neither matrix commutes with ${\cal V}^{(m)}$. 

For later use, we note that ${\cal J}$ has eigenvalues
\begin{equation}
j_{l}=2\cos^{2}(\pi l/N),\;\;l=1,2,\ldots N,\label{jldef}
\end{equation}
corresponding to the eigenvectors $\psi^{(l)}$ with elements
\begin{equation}
\psi^{(l)}_{n}=N^{-1/2}\exp(2\pi iln/N).\label{psildef} 
\end{equation}
The eigenvalues of ${\cal K}$ are
\begin{equation}
\kappa_{l}=i\sin(2\pi l/N),\;\;l=1,2,\ldots N,\label{kappaldef}
\end{equation}
for the same eigenvectors $\psi^{(l)}$. From Eq.\ \eqref{jldef} we see that for $N$ even there is a zero eigenvalue of ${\cal J}$ (at $l=N/2$). To avoid the complications from a noninvertible ${\cal J}$, we restrict ourselves to $N$ odd (when all eigenvalues of ${\cal J}$ are nonzero).

\subsection{Transfer matrix}
\label{transfer}

The discretized Dirac equation is expressed in terms of the matrices \eqref{Idef}--\eqref{Vdef} by
\begin{align}
\frac{1}{2\Delta}{\cal J}(\bm{\Psi}_{m+1}-\bm{\Psi}_{m})={}&\left(-\frac{i}{2\Delta}\sigma_{z}{\cal K}-\frac{i}{4}\sigma_{x}{\cal V}^{(m)}\right)\nonumber\\
&(\bm{\Psi}_{m}+\bm{\Psi}_{m+1}).\label{Diracdiscrete}
\end{align}
Rearranging Eq.\ \eqref{Diracdiscrete} we arrive at Eq.\ \eqref{transfermatrix} with the transfer matrix
\begin{align}
{\cal M}_{m}={}&\left({\cal J}+i\sigma_{z}{\cal K}+\tfrac{1}{2}i\Delta\sigma_{x}{\cal V}^{(m)}\right)^{-1}\nonumber\\
&\left({\cal J}-i\sigma_{z}{\cal K}-\tfrac{1}{2}i\Delta\sigma_{x}{\cal V}^{(m)}\right).\label{Mresult}
\end{align}
Since we take $N$ odd, so that ${\cal J}$ is invertible, we may equivalently write Eq.\ \eqref{Mresult} in the more compact form
\begin{equation}
{\cal M}_{m}=\frac{1-iX_{m}}{1+iX_{m}},\;\; X_{m}={\cal J}^{-1}(\sigma_{z}{\cal K}+\tfrac{1}{2}\Delta\sigma_{x}{\cal V}^{(m)}).\label{MXrelation}
\end{equation}
As announced, the transfer matrix is nonlocal (in the sense that multiplication of $\bm{\Psi}_{m}$ by ${\cal M}_{m}$ couples all transverse coordinates).

One can readily check that the condition \eqref{Msymplectic} of symplectic symmetry is fullfilled. In App.\ \ref{App_current} we demonstrate that the condition \eqref{currentconservation} of current conservation holds if we define the discretized current operator $J_{x}$ in terms of the symmetric matrix ${\cal J}$,
\begin{equation}
J_{x}=\tfrac{1}{2}v\sigma_{x}{\cal J}.\label{Jxdef}
\end{equation}
The absence of fermion doubling is checked in Sec.\ \ref{dispersion}.

The transfer matrix ${\cal M}$ through the entire strip (from $x=0$ to $x=L$) is the product of the one-step transfer matrices ${\cal M}_{m}$,
\begin{equation}
{\cal M}=\prod_{m=1}^{M}{\cal M}_{m},\label{MMmrelation}
\end{equation}
ordered such that ${\cal M}_{m+1}$ is to the left of ${\cal M}_{m}$. The properties of symplectic symmetry and current conservation are preserved upon matrix multiplication.

\subsection{Numerical stability}
\label{stability}

The repeated multiplication \eqref{MMmrelation} of the one-step transfer matrix to arrive at the transfer matrix of the entire strip is unstable because it produces both exponentially growing and exponentially decaying eigenvalues, and the limited numerical accuracy prevents one from retaining both sets of eigenvalues. We resolve this obstacle, following Refs.\ \cite{Bar07,Sny08,Tam91}, by converting the transfer matrix into a unitary matrix, which has only eigenvalues of unit absolute value. The formulas that accomplish this transformation are given in App.\ \ref{Mmultiply}.

\section{From transfer matrix to scattering matrix and conductance}
\label{fromMtoS}

\subsection{General formulation}
\label{general}

The scattering matrix is obtained from the transfer matrix by connecting the two ends of the strip at $x=0$ and $x=L$ to semi-infinite ballistic leads. The $N$ transverse modes in the leads (calculated in Sec.\ \ref{ballistic}), consist of $N_{0}$ propagating modes $\phi_{l}^{\pm}$ (labeled $+$ for right-moving and $-$ for left-moving), and $N-N_{0}$ evanescent modes $\chi_{l}^{\pm}$ (decaying for $x\rightarrow\pm\infty$). The propagating modes are normalized such that each carries unit current.

Consider an incoming wave in mode $l_{0}$ from the left. At $x=0$, the sum of incoming, reflected, and evanescent waves is given by
\begin{equation}
\Phi^{\rm left}_{l_{0}}=\phi_{l_{0}}^{+}+\sum_{l}r_{l,l_{0}}\phi_{l}^{-}+\sum_{l}\alpha_{l,l_{0}}\chi_{l}^{-},\label{Phileft}
\end{equation}
while the sum of transmitted and evanescent waves at $x=L$ is given by
\begin{equation}
\Phi^{\rm right}_{l_{0}}=\sum_{l}t_{l,l_{0}}\phi_{l}^{+}+\sum_{l}\alpha'_{l,l_{0}}\chi_{l}^{+}.\label{Phiright}
\end{equation}

The $N_{0}\times N_{0}$ reflection matrix $r$ and transmission matrix $t$ are obtained by equating
\begin{equation}
\Phi^{\rm right}_{l_{0}}={\cal M}\Phi^{\rm left}_{l_{0}},\label{PhiM}
\end{equation}
eliminating the coefficients $\alpha,\alpha'$, and repeating for each of the $N_{0}$ propagating modes incident from the left. Starting from a mode incident from the right, we similarly obtain the reflection matrices $r'$ and $t'$, which together with $r$ and $t$ form a $2N_{0}\times 2N_{0}$ unitary scattering matrix,
\begin{equation}
S=\begin{pmatrix}
r&t'\\
t&r'
\end{pmatrix}.\label{Sdef}
\end{equation}
As a consequence of unitarity, the matrix products $tt^{\dagger}$ and $t't'^{\dagger}$ have the same set of eigenvalues $T_{1},T_{2},\ldots T_{N_{0}}$, called transmission eigenvalues.

The number $N_{0}$ of propagating modes in the leads is an odd integer, because of our choice of periodic boundary conditions. The symplectic symmetry condition \eqref{Msymplectic} then implies that the transmission eigenvalues $T_{n}$ consist of one unit eigenvalue and $(N_{0}-1)/2$ degenerate pairs (Kramers degeneracy \cite{note1}).

The conductance $G$ follows from the transmission eigenvalues via the Landauer formula,
\begin{equation}
G=G_{0}\sum_{n}T_{n}.\label{GG0}
\end{equation}
The conductance quantum $G_{0}=4e^{2}/h$ in the application to graphene (which has both spin and valley degeneracies), while $G_{0}=e^{2}/h$ in the application to the surface of a topological insulator. The Kramers degeneracy, which is present in both applications, is accounted for in the sum over the transmission eigenvalues.

\subsection{Infinite wave vector limit}
\label{infinitelimit}

Following Ref.\ \cite{Two06}, we model metal contacts by leads with an infinitely large Fermi wave vector. In the infinite wave vector limit all modes in the leads are propagating, so $N_{0}=N$ and the scattering matrix has dimension $2N\times 2N$. The states $\phi_{l}^{\pm}$ ($l=1,2,\ldots N$) in this limit are simply the $2N$ eigenstates of the current operator $J_{x}$, normalized such that each carries the same current. In terms of the eigenvalues and eigenvectors \eqref{jldef}, \eqref{psildef} of ${\cal J}$ we have
\begin{equation}
\phi_{l}^{\pm}=j_{l}^{-1/2}\begin{pmatrix}
1\\ \pm 1
\end{pmatrix}
\psi^{(l)}.\label{philinfinite}
\end{equation}

Instead of the general Eqs.\ \eqref{Phileft} and \eqref{Phiright} we now have the simpler equations
\begin{equation}
\Phi^{\rm left}_{l_{0}}=\phi_{l_{0}}^{+}+\sum_{l}r_{l,l_{0}}\phi_{l}^{-},\;\;
\Phi^{\rm right}_{l_{0}}=\sum_{l}t_{l,l_{0}}\phi_{l}^{+}.\label{Phileftright}
\end{equation}
To obtain from Eq.\ \eqref{PhiM} a closed-form expression for $S$ in terms of ${\cal M}$, we first perform the similarity transformation
\begin{equation}
\tilde{\cal M}={\cal R}{\cal M}{\cal R}^{-1},\;\;{\cal R}=\sigma_{H}{\cal J}^{1/2},\label{tildeM}
\end{equation}
where $\sigma_{H}$ is the Hadamard matrix,
\begin{equation}
\sigma_{H}=2^{-1/2}\begin{pmatrix}
1&1\\
1&-1
\end{pmatrix}=\sigma_{H}^{-1}.\label{Hdef}
\end{equation}
The notation $\sigma_{H}{\cal J}^{1/2}$ signifies a direct product, where $\sigma_{H}$ acts on the spinor degrees of freedom $s=\pm$ and ${\cal J}^{1/2}$ acts on the lattice degrees of freedom $n=1,2,\ldots N$. Notice that the matrix ${\cal R}$ is Hermitian [since ${\cal J}$ is Hermitian with exclusively positive eigenvalues, see Eq.\ \eqref{jldef}].

We separate the spinor degrees of freedom of ${\cal M}$ into four $N\times N$ blocks,
\begin{equation}
{\cal M}=\begin{pmatrix}
{\cal M}^{++}&{\cal M}^{+-}\\
{\cal M}^{-+}&{\cal M}^{--}
\end{pmatrix},\label{Msubblock}
\end{equation}
such that ${\cal M}_{ns,ms'}={\cal M}_{nm}^{ss'}$. The matrix $\tilde{\cal M}$ has a corresponding decomposition into submatrices $\tilde{\cal M}^{ss'}$. As one can verify by substitution into Eq.\ \eqref{Phileftright} and comparison with Eq.\ \eqref{PhiM}, the submatrices $\tilde{\cal M}^{ss'}$ are related to the transmission and reflection matrices by
\begin{subequations}
\label{ttos}
\begin{eqnarray}
r&=&-\left(\tilde{\cal M}^{--}\right)^{-1} \tilde{\cal M}^{-+},\label{ttosa}\\
t&=&\tilde{\cal M}^{++}-\tilde{\cal M}^{+-}\left(\tilde{\cal M}^{--}\right)^{-1} \tilde{\cal M}^{-+},\label{ttosb}\\
t'&=&\left(\tilde{\cal M}^{--}\right)^{-1},\label{totsc}\\
r'&=&\tilde{\cal M}^{+-}\left(\tilde{\cal M}^{--}\right)^{-1}.\label{ttosd}
\end{eqnarray}
\end{subequations}

Similar formulas were derived in Ref.\ \cite{Sny08}, but there the transformation from ${\cal M}$ to $S$ involved only a Hadamard matrix and no matrix ${\cal J}$, because of the different current operator in that model.

\section{Ballistic transport}
\label{ballistic}

For a constant $U$ we have ballistic transport through the strip of length $L$ and width $W$. In this section we check that we recover the known results \cite{Kat06,Two06} for ballistic transport of Dirac fermions from the discretized transfer matrix.

\subsection{Dispersion relation}
\label{dispersion}

For $U=U_{0}=\mbox{constant}$ the matrix \eqref{Vdef} of discretized potentials is given by ${\cal V}^{(m)}=-(\varepsilon/\Delta){\cal J}$, with $\varepsilon =(E-U_{0})\Delta/\hbar v$ the dimensionless energy (measured relative to the Dirac point at energy $U_{0}$). Substitution into Eq.\ \eqref{MXrelation} gives the $m$-independent ballistic transfer matrix ${\cal M}_{\rm ball}$,
\begin{equation}
{\cal M}_{\rm ball}=\frac{1+i(\varepsilon/2)\sigma_{x}-i{\cal J}^{-1}{\cal K}\sigma_{z}}{1-i(\varepsilon/2)\sigma_{x}+i{\cal J}^{-1}{\cal K}\sigma_{z}}.\label{Mball}
\end{equation}
This is the one-step transfer matrix. The transfer matrix through the entire strip, in this ballistic case, is simply ${\cal M}=({\cal M}_{\rm ball})^{M}$.

In accordance with Eqs.\ \eqref{jldef}--\eqref{kappaldef}, the matrix ${\cal J}^{-1}{\cal K}$ can be diagonalized (for $N$ odd) by
\begin{subequations}
\label{JKLambdaF}
\begin{align}
&{\cal J}^{-1}{\cal K}={\cal F}\Lambda{\cal F}^{\dagger},\;\;\Lambda_{nn'}=i\tan(\pi n/N)\delta_{nn'},\label{JKLambdaFa}\\
&{\cal F}_{nn'}=N^{-1/2}\exp(2\pi i nn'/N).\label{JKLambdaFb}
\end{align}
\end{subequations}
The Fourier transformed transfer matrix ${\cal F}{\cal M}_{\rm ball}{\cal F}^{\dagger}$ is diagonal in the mode index $l=1,2,\ldots N$. A $2\times 2$ matrix structure $m_{l}$ in the spin index remains, given by
\begin{equation}
m_{l}=\frac{1+i(\varepsilon/2)\sigma_{x}+\tan(\pi l/N)\sigma_{z}}{1-i(\varepsilon/2)\sigma_{x}-\tan(\pi l/N)\sigma_{z}}.\label{mldef}
\end{equation}
The eigenvalues and eigenvectors of $m_{l}$ are
\begin{equation}
m_{l} u_{l}^{\pm}=e^{\pm ik} u_{l}^{\pm},\;\;
 u_{l}^{\pm}=\begin{pmatrix}
\varepsilon/2\\
i\tan(\pi l/N)\pm\tan(k/2)
\end{pmatrix},\label{mlu}
\end{equation}
with the dimensionless momentum $k$ given as a function of $\varepsilon$ and $l$ by the dispersion relation
\begin{equation}
\tan^{2}(k/2)+\tan^{2}(\pi l/N)=(\varepsilon/2)^{2}.\label{dispersionrelation}
\end{equation}

\begin{figure}[tb]
\centerline{\includegraphics[width=0.9\linewidth]{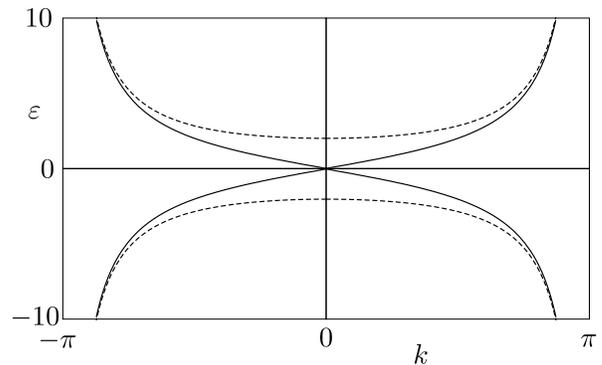}}
\caption{\label{fig_dispersion}
Dispersion relation \eqref{dispersionrelation} of the discretized Dirac equation, plotted in the first Brillouin zone for two transverse modes ($l=N$, solid curve; $l\approx N/4$, dotted curve). The dispersion relation approaches that of the Dirac equation near the point $(k,\varepsilon)=(0,0)$, and avoids fermion doubling at other points in the Brillouin zone.}
\end{figure}

In Fig.\ \ref{fig_dispersion} we have plotted the dispersion relation \eqref{dispersionrelation} for two different modes in the first Brillouin zone $-\pi<k<\pi$. For each mode index $l$, there is one wave that propagates to positive $x$ (on the branch with $d\varepsilon/dk>0$) and one wave that propagates to negative $x$ (on the branch with $d\varepsilon/dk<0$).

As anticipated \cite{Sta82,Ben83}, the discretization of the Dirac equation on the displaced lattice (crosses in Fig.\ \ref{fig_lattice}) has avoided the spurious doubling of the fermion degrees of freedom that would have happened if the finite differences would have been calculated on the original lattice (solid dots in Fig.\ \ref{fig_lattice}). In the low-energy and long-wave-length limit $k,\varepsilon\rightarrow 0$, the conical dispersion relation $(vp_{x})^{2}+(vp_{y})^{2}=(E-U_{0})^{2}$ of the Dirac equation \eqref{Dirac} is recovered. The longitudinal momentum is $p_{x}=\hbar k/\Delta$, while the transverse momentum is $p_{y}=(2\pi\hbar/W)l$ if $l/N\rightarrow 0$ or $p_{y}=-(2\pi\hbar/W)(N-l)$ if $l/N\rightarrow 1$.

\subsection{Evanescent modes}
\label{evanescent}

For $|\varepsilon|<2\tan(\pi/N)$, hence for $|E-U_{0}|\lesssim 2\pi\hbar v/W$, only the mode with index $l=N$ is propagating. The other $N-1$ modes are evanescent, that is to say, their wave number $k$ has a nonzero imaginary part $\kappa$. There are two classes of evanescent modes, one class with a purely imaginary wave number $k=i\kappa_{+}$, and another class with a complex wave number $k=\pi+i\kappa_{-}$. The relation between $\kappa_{\pm}$ and $\varepsilon$, following from Eq.\ \eqref{dispersionrelation}, is
\begin{subequations}
\label{kappadef}
\begin{align}
&\mbox{tanh}\,(\kappa_{+}/2)=\tan^{2}(\pi l/N)-(\varepsilon/2)^{2},\label{kappaplus}\\
&\mbox{cotanh}\,(\kappa_{-}/2)=\tan^{2}(\pi l/N)-(\varepsilon/2)^{2}.\label{kappamin}
\end{align}
\end{subequations}

\begin{figure}[tb]
\centerline{\includegraphics[width=0.8\linewidth]{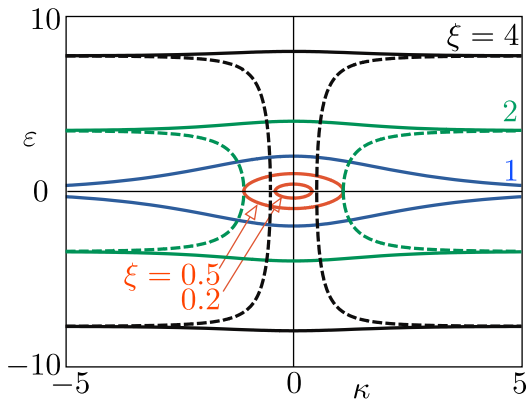}}
\caption{\label{fig_evanescent}
Relation between the energy $\varepsilon$ and the imaginary part $\kappa$ of the wave number of evanescent modes, calculated from Eq.\ \eqref{kappadef} for five different values of the mode index [parameterized by $\xi=\tan(\pi l/N)$]. The real part of the wave number equals $0$ on the solid contours (corresponding to $\kappa_{+}$), while it equals $\pi$ on the dashed contours (corresponding to $\kappa_{-}$). Only the $\kappa_{+}$ evanescent modes have a correspondence to the Dirac equation in the limit $\varepsilon\rightarrow 0$. The $\kappa_{-}$ evanescent modes that appear for $|\xi|>1$ are artefacts of the discretization for large transverse momenta.
}
\end{figure}

In Fig.\ \ref{fig_evanescent} we have plotted Eq.\ \eqref{kappadef} for different mode indices, parameterized by $\xi=\tan(\pi l/N)$. The evanescent modes in the Dirac equation correspond to $k=i\kappa_{+}$ in the limit $\varepsilon\rightarrow 0$ (solid contours in Fig.\ \ref{fig_evanescent}). The second ``spurious'' class of evanescent modes, with $k=\pi+i\kappa_{-}$ (dashed contours), is an artefact of the discretization that appears for large transverse momenta ($|\xi|>1$, or $N/4< l< 3N/4$).

To minimize the effect of the spurious evanescent modes we insert a pair of filters of length $L_{0}$ between the strip of length $L$ and the leads with infinitely large Fermi wave vector. By choosing a large but finite Fermi wave vector in the filters, they remove the spurious evanescent modes of large transverse momenta which are excited by the infinite Fermi wave vector in the leads.

\begin{figure}[tb]
\centerline{\includegraphics[width=0.8\linewidth]{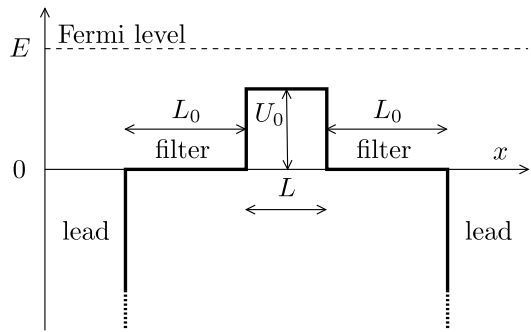}}
\caption{\label{fig_geometry}
Potential profile of a strip (length $L$), connected to leads by a pair of filters (length $L_{0}$). The Fermi wave vector in the leads is taken infinitely large; the finite Fermi wave vector in the filters removes the spurious evanescent modes excited by the leads.
}
\end{figure}

The geometry is sketched in Fig.\ \ref{fig_geometry}. In the filters we choose $U=0$, $E=2\hbar v/\Delta$ (so $\varepsilon=2$ in the filters). Since $|\kappa_{-}|\geq\pi/N$  for the spurious evanescent modes [described by Eq.\ \eqref{kappamin}], their longest decay length is of order $N\Delta=W$. By choosing $L_{0}=10\,W$ we ensure that these modes are filtered out. 

\subsection{Conductance}
\label{conductance}

\begin{figure}[tb]
\centerline{\includegraphics[width=0.9\linewidth]{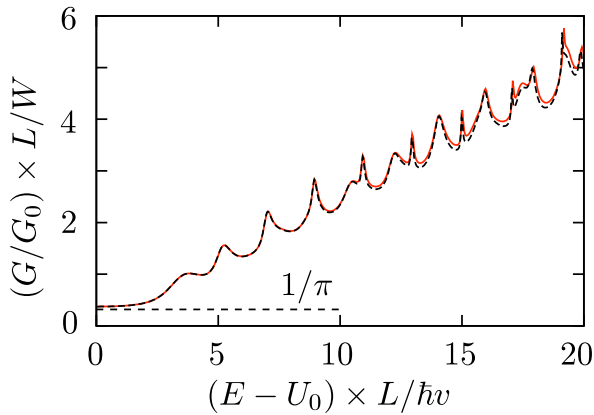}}
\caption{\label{fig_ballisticG}
Solid curve: conductance in the geometry of Fig.\ \ref{fig_geometry}. The Fermi
wave vector $(E-U_{0})/\hbar v$ in the strip of length $L$ and width $W=3L$ is
varied by varying the potential step height $U_{0}$ at fixed Fermi energy
$E=2\hbar v/\Delta$. The lattice constant $\Delta=10^{-2}\,L$. Dashed curve: the
result from the Dirac equation (calculated from the formulas in Ref.\
\cite{Two06}), corresponding to the limit $\Delta\rightarrow 0$. The horizontal dotted line is the mimimal conductivity at the Dirac point.
}
\end{figure}

We have calculated the conductance at fixed Fermi energy $E=2\hbar v/\Delta$ as a function of the potential step height $U_{0}$. Results are shown in Fig.\ \ref{fig_ballisticG} for aspect ratio $W/L=3$ and lattice constant $\Delta=10^{-2}\,L$ (solid curve) and compared with the solution of the Dirac equation (dashed curve). The agreement is excellent (for a twice smaller $\Delta$ the two curves would have been indistinguishable).

The horizontal dotted line in Fig.\ \ref{fig_ballisticG} indicates the value \cite{Kat06,Two06}
\begin{equation}
\lim_{W/L\rightarrow\infty}\;\lim_{U_{0}\rightarrow E}(L/W)G/G_{0}= 1/\pi \label{Gminimal}
\end{equation}
of the minimal conductivity at the Dirac point for a large aspect ratio of the strip. The oscillations which develop as one moves away from the Dirac point are Fabry-Perot resonances from multiple reflections at $x=0$ and $x=L$. The filters of length $L_{0}$ are not present in the continuum calculation (dashed curve), but the close agreement with the lattice calculation (solid curve) shows that the filters do not modify these resonances in any noticeable way. The filters do play an essential role in ensuring that the minimal conductivity reaches its proper value \eqref{Gminimal}: Without the filters the lattice calculation would give a twice larger minimal conductivity, due to the contribution from the spurious evanescent modes of large transverse momentum.

\section{Transport through disorder}
\label{disorder}

We introduce disorder in the strip of length $L$ by adding a random potential $\delta U$ to each lattice point, distributed uniformly in the interval $(-\Delta U,\Delta U)$. Since our discretization scheme conserves the symplectic symmetry exactly, there is no need now to choose a finite correlation length for the potential fluctuations (as in earlier numerical studies \cite{Bar07,Nom07,Ryc07,Lew08,Zha08,Sny08,San07,Nom08}). Instead we can let the potential of each lattice point fluctuate independently, as in the original Anderson model of localization \cite{And58}.

\subsection{Scaling of conductance at the Dirac point}
\label{scaling}

\begin{figure}[tb]
\centerline{\includegraphics[width=0.8\linewidth]{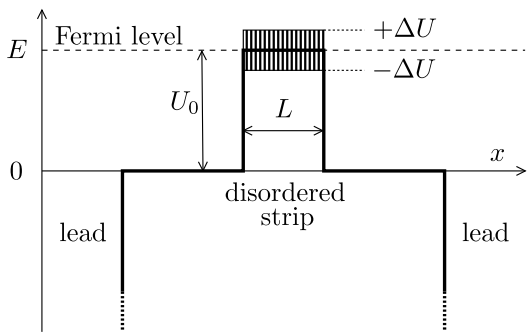}}
\caption{\label{fig_geometry_D}
Same as Fig.\ \ref{fig_geometry}, but now for the case that the potential in the strip fluctuates around the Dirac point: $U=U_{0}+\delta U$, with $U_{0}=E$ and $\delta U$ uniformly distributed in the interval $(-\Delta U,\Delta U)$.
}
\end{figure}

\begin{figure}[tb]
\centerline{\includegraphics[width=0.9\linewidth]{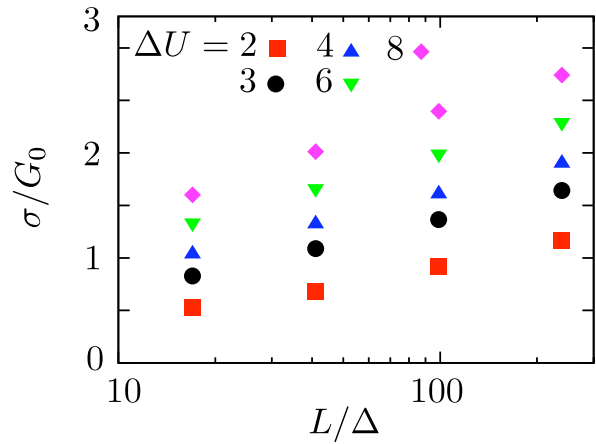}}
\caption{\label{figscale1}
Scaling with system size of the average conductivity $
\sigma\equiv (L/W)\langle G\rangle$ in a disordered strip at the Dirac point (geometry of Fig.\ \ref{fig_geometry_D}). The length $L$ of the strip  is varied at fixed aspect ratio $W/L=3$. The data are collected for different disorder strengths $\Delta U$ (listed in units of $\hbar v/\Delta$).
} 
\end{figure}

When $U_{0}=E$ the potential $U_{0}+\delta U$ in the strip fluctuates around the Dirac point (see Fig.\ \ref{fig_geometry_D}). Results for the scaling of the average conductivity $\sigma\equiv(L/W)\langle G\rangle$ with system size are shown for different disorder strengths in Fig.\ \ref{figscale1}. We averaged over 3000 disorder realizations for $L/\Delta=17,41,99$ and over 300 realizations for $L/\Delta=239$. The aspect ratio was fixed at $W/L=3$.

\begin{figure}[tb]
\centerline{\includegraphics[width=0.9\linewidth]{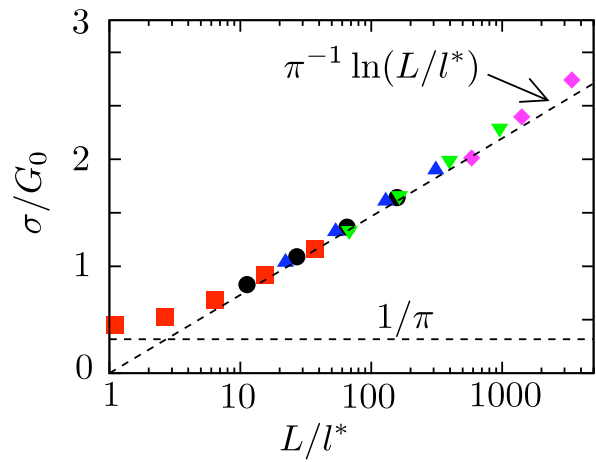}}
\caption{\label{figscale2}
Dependence of the conductivity of Fig.\ \ref{figscale1} on the rescaled system length $L/l^{\ast}(\Delta U)$. The two dotted lines are the analytical weak and strong disorder limits. 
}
\end{figure}

For sufficiently strong disorder strengths $\Delta U \gtrsim 3\hbar v/\Delta$ the data follow the logarithmic scaling \cite{Bar07,Nom07}
\begin{equation}
\sigma/G_{0} = c \ln[L/l^{\ast}(\Delta U)].\label{sigmalog}
\end{equation} 
There is a consensus in the literature that $c=1/\pi$  can be calculated perturbatively \cite{Sch08} as a weak antilocalization correction. The quantity $l^{\ast}$ plays the role of a mean free path, dependent on the disorder strength. We fit this scaling to our data with a common fitting parameter $c$ (disregarding the data sets with low $\Delta U$ as being too close to the ballistic limit). The fitting gives $l^{\ast}$ for every data set with the same $\Delta U$. 

The resulting single-parameter scaling is presented in Fig.\ \ref{figscale2} (including also the low $\Delta U$ sets, for completeness). The data sets collapse onto a single logarithmically increasing conductivity with $c\approx 0.33(1)$, close to the expected value $c=1/\pi\approx 0.318$. To assess the importance of finite-size corrections \cite{Sle04} we include a non-universal lattice-constant dependent term to the logarithmic scaling: $\sigma/G_{0}=c\ln[L/l^{\ast}(\Delta U)]+f(\Delta U)\Delta/L$. We then find $c\approx 0.316(5)$, again close to the expected value \cite{Sch08}. These results for the absence of localization of Dirac fermions are consistent with earlier numerical calculations \cite{Bar07,Nom07} using a momentum space regularization of the Dirac equation.

\subsection{Conductance fluctuations at the Dirac point}
\label{UCF}

The sample-to-sample conductance fluctuations at the Dirac point were calculated numerically in Ref.\ \cite{Ryc07} using the tight-binding model on a honeycomb lattice. An enhancement of the variance above the value for point scatterers was observed, and explained in Ref.\ \cite{Kha08} in terms of the absence of intervalley scattering. A perturbative calculation \cite{Kha08,Kec08} of ${\rm Var}\,G=\langle G^{2}\rangle-\langle G\rangle^{2}$ gives
\begin{equation}
{\rm Var}\,G=\frac{3\zeta(3)}{\pi^{3}}\,\frac{W}{L}\,G_{0}^{2},\;\;W/L\gg 1.\label{varG}
\end{equation}
Intervalley scattering would reduce the variance by a factor of four, while trigonal warping without intervalley scattering would reduce the variance by a factor of two.

\begin{figure}[tb]
\centerline{\includegraphics[width=0.9\linewidth]{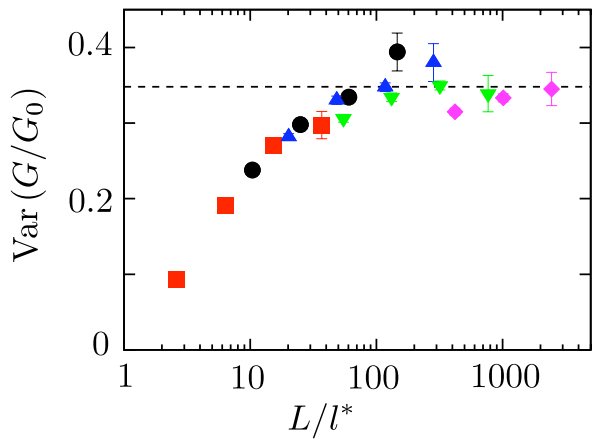}}
\caption{\label{figvar}
Same as Fig.\ \ref{figscale2}, but now for the variance of the conductance (instead of the ensemble average). The horizontal dotted line is the analytical prediction \eqref{varG}, ${\rm Var}\,(G/G_{0})=0.116\,W/L$ with $W/L=3$. 
}
\end{figure}

In Fig.\ \ref{figvar} we plot our results for the dependence of the variance of the conductance on the rescaled system size $L/l^{\ast}$, with the $\Delta U$ dependence of $l^{\ast}$ obtained from the scaling analysis of the average conductance in Sec.\ \ref{scaling}. The convergence towards the expected value \eqref{varG} is apparent. The numerical data of Fig.\ \ref{figvar} supports the conclusion of Ref.\ \cite{Sch08}, that the statistics of the conductance at the Dirac point can be obtained from metallic diffusive perturbation theory in the large-$L$ limit.

The tight-binding model calculation of Ref.\ \cite{Ryc07} only reached about half the expected value \eqref{varG}, presumably because the potential was not quite smooth enough to avoid intervalley scattering. This illustrates the power of the finite difference method used here: We retain single-valley physics even when the correlation length of the potential is equal to the lattice constant. 

\subsection{Transport away from the Dirac point}
\label{away}

\begin{figure}[tb]
\centerline{\includegraphics[width=0.9\linewidth]{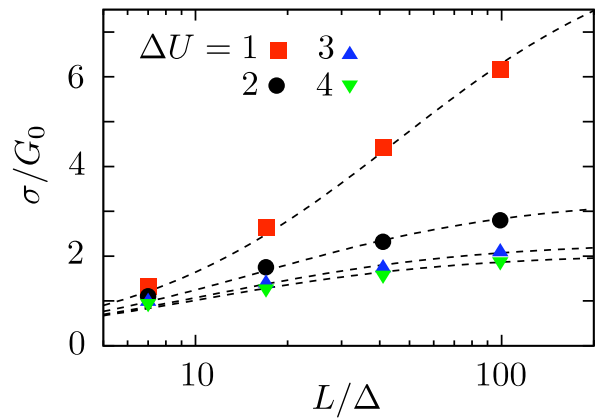}}
\caption{\label{figdiff}
Crossover from ballistic to diffusive conduction away from the Dirac point. The conductivity is plotted versus system size, at fixed Fermi wave vector $(E-U_{0})/\hbar v=0.8\,\Delta^{-1}$ in the strip and fixed aspect ratio $W/L=3$. The data is for different disorder strengths $\Delta U$, listed in units of $\hbar v/\Delta$. The dotted curves are a fit to the semiclassical formula \eqref{mfp}, with the transport mean free path $l_{0}$ as a fit parameter.
}
\end{figure}

\begin{figure}[tGb]
\centerline{\includegraphics[width=0.9\linewidth]{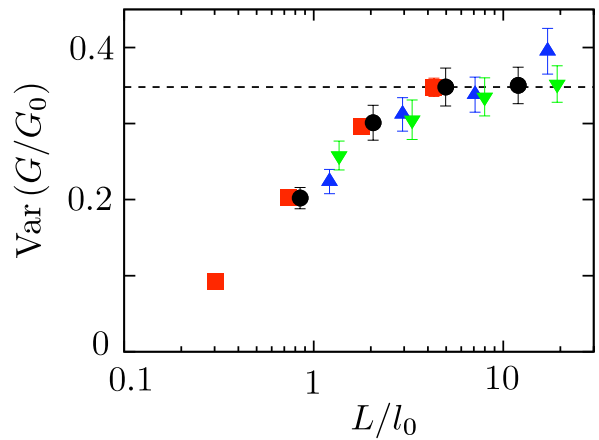}}
\caption{\label{figvardiff}
Same as Fig.\ \ref{figdiff}, but now for the variance of the conductance. The data is plotted as a function of the rescaled sample size, using the values of the mean free path obtained from the fit of the conductance. The horizontal dotted line is the analytical prediction \eqref{varG}.
}
\end{figure}

The results of Secs.\ \ref{scaling} and \ref{UCF} are for potential fluctuations around the Dirac point ($U_{0}=E$). In this subsection we consider the average conductance and the conductance fluctuations away from the Dirac point. We take $(E-U_{0})=0.8\,\hbar v/\Delta$ and vary the sample length $L$ at fixed aspect ratio $W/L=3$. The resulting size dependence of the conductivity is presented in Fig.\ \ref{figdiff}, for different disorder strengths $\Delta U$. 

Since antilocalization is a relatively small quantum correction at these high Fermi energies, we are in the regime described by the semiclassical Boltzmann equation \cite{Ada07,Ada08}. In App.\ \ref{crossover} we apply a general theory \cite{Paa08} for the crossover from ballistic to diffusive conduction, to arrive at the formula
\begin{equation}
\langle G\rangle=\frac{\pi}{2}G_{0}N_{\rm strip}\frac{l_{0}}{L+2l_{0}},\label{mfp}
\end{equation}
for the average conductance in terms of the transport mean free path $l_{0}$ and the number $N_{\rm strip}=|E-U_{0}|(W/\pi\hbar v)$ of propagating modes in the strip. From the fit of $\langle G\rangle$ versus $L$ in Fig.\ \ref{figdiff} we extract the dependence on $\Delta U$ of $l_{0}$, and then we use that information to investigate the scaling of the variance of the conductance with system size. As seen in Fig.\ \ref{figvardiff}, the variance scales well towards the expected value \eqref{varG}.

\section{Conclusion}
\label{conclude}

In conclusion, we have presented in this paper what one might call the ``Anderson model for Dirac fermions''. Just as in the original Anderson tight-binding model of localization \cite{And58}, our model is a tight-binding model on a lattice with uncorrelated on-site disorder. Unlike the tight-binding model of graphene (with nearest neighbor hopping on a honeycomb lattice), our model preserves the symplectic symmetry of the Dirac equation --- at the expense of a nonlocal finite difference approximation of the transfer matrix.

Our finite difference method is based on a discretization scheme developed in the context of lattice gauge theory \cite{Sta82,Ben83}, with the purpose of resolving the fermion doubling problem. We have adapted this scheme to include the chiral symmetry breaking by a disorder potential, and have cast it in a current-conserving transfer matrix form suitable for the calculation of transport properties.

To test the validity and efficiency of the model, we have calculated the average and the variance of the conductance and compared with earlier numerical and analytical results. We recover the logarithmic increase of the average conductance at the Dirac point, found in numerical calculations that use a momentum space rather than a real space discretization of the Dirac equation \cite{Bar07,Nom07}. The coefficient that multiplies the logarithm is close to $1/\pi$, in agreement with analytical expectations \cite{Sch08}. The variance of the conductance is enhanced by the absence of intervalley scattering, and we have been able to confirm the scaling with increasing system size towards the expected limit \cite{Kha08,Kec08} --- something which had not been possible in earlier numerical calculations \cite{Ryc07} because intervalley scattering sets in before the large-system limit is reached.

\begin{figure}[tGb]
\centerline{\includegraphics[width=0.9\linewidth]{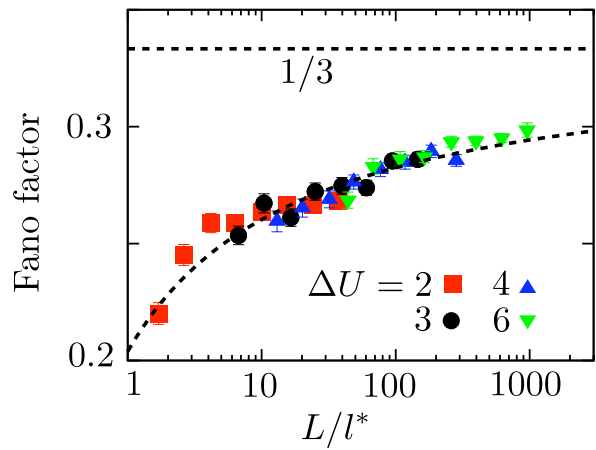}}
\caption{\label{scalefano}
Scaling with system size of the Fano factor (average shot noise power divided by average current) in a disordered strip at the Dirac point (geometry of Fig.\ \ref{fig_geometry_D}). The length $L$ of the strip  is varied at fixed aspect ratio $W/L=3$. The data are collected for different disorder strengths $\Delta U$ (listed in units of $\hbar v/\Delta$). The dotted horizontal line is the value $F=1/3$ for a diffusive metal. The dotted curve is a fit to $F=1/3+a[b+\ln(L/l^{\ast})]^{-1}$, included in order to indicate a possible scaling towards the expected value.
}
\end{figure}

Our calculations support the expectation \cite{Sch08} that the statistics of the conductance at the Dirac point scales towards that of a diffusive metal in the large-system limit. This would imply that the shot noise should scale towards a Fano factor $F=1/3$ \cite{Bee92}. Earlier numerical studies using the momentum space discretization \cite{San07} found a saturation at the smaller value of $F=0.295$. Our own numerical results, shown in Fig.\ \ref{scalefano}, instead suggest a slow, logarithmic, increase towards the expected $F=1/3$. More research on this particular quantity is required for a conclusive answer.

We anticipate that the numerical method developed here will prove useful for the study of graphene with smooth disorder potentials (produced for example by remote charge fluctuations), since such potentials produce little intervalley scattering. Intervalley scattering is absent by construction in the metallic surface states of topological insulators (such as BiSb \cite{Hsi08}). These surface states might be studied by starting from a three-dimensional tight-binding model, but we would expect a two-dimensional formulation as presented here to be more efficient.

\acknowledgments
We have benefited from the insight of A. R. Akhmerov, I. Snyman, and S. V. Syzranov. This research was supported by the Dutch Science Foundation NWO/FOM.

\appendix

\section{Current conserving discretization of the current operator}
\label{App_current}

We seek a discretization of the current operator \eqref{jdef} that satisfies the condition \eqref{currentconservation} of current conservation. Substitution of the expression \eqref{MXrelation} into the condition \eqref{currentconservation} gives the requirement
\begin{equation}
J_{x}^{-1}{\cal M}_{m}^{\dagger}J_{x}={\cal M}_{m}^{-1}\Leftrightarrow
J_{x}^{-1}X_{m}^{\dagger}J_{x}=X_{m}.\label{JMrequirement}
\end{equation}
The requirement that Eq.\ \eqref{JMrequirement} holds for any choice of potential fixes the discretization \eqref{Jxdef} of the current operator [up to a multiplicative constant, which follows from the continuum limit \eqref{jdef}].

This is an appropriate point to note that current conservation could not have been achieved if the potential would have been discretized in a way that would have resulted in a nonsymmetric matrix ${\cal V}_{m}$. For example, if instead of Eq.\ \eqref{VPsi} we would have chosen
\begin{align}
V\Psi\rightarrow{}&\tfrac{1}{4}(\tilde{V}_{m+1,n}\Psi_{m+1,n}+\tilde{V}_{m+1,n+1}\Psi_{m+1,n+1}\nonumber\\
&+\tilde{V}_{m,n}\Psi_{m,n}+\tilde{V}_{m,n+1}\Psi_{m,n+1}),\label{VPsiwrong}
\end{align}
with $\tilde{V}_{m,n}=V(x_{m},y_{n})$, then the corresponding matrix ${\cal V}_{m}$ would have been asymmetric and no choice of $J_{x}$ could have satisfied Eq.\ \eqref{JMrequirement}.

\section{Stable multiplication of transfer matrices}
\label{Mmultiply}

To perform the multiplication \eqref{MMmrelation} of transfer matrices in a stable way (avoiding exponentially growing and decaying eigenvalues), we use the current conservation relation \eqref{currentconservation} to convert the product into a composition of unitary matrices (involving only eigenvalues of unit absolute value). The same method was used in Refs.\ \cite{Bar07,Sny08,Tam91}, but for a different current operator, so the required transformation formulas need to be adapted.

We separate the spinor degrees of freedom $s=\pm$ of the transfer matrix ${\cal M}_{m}$ into four $N\times N$ blocks,
\begin{equation}
{\cal M}_{m}=\begin{pmatrix}
{\cal M}_{m}^{++}&{\cal M}_{m}^{+-}\\
{\cal M}_{m}^{-+}&{\cal M}_{m}^{--}
\end{pmatrix}.\label{MsubblockB}
\end{equation}
The current conservation relation \eqref{currentconservation} with current operator \eqref{Jxdef} can be written in the canonical form,
\begin{equation}
\tilde{\cal M}^{\dagger}_{m}
\begin{pmatrix}
1&0\\
0&-1
\end{pmatrix}
\tilde{\cal M}_{m}=
\begin{pmatrix}
1&0\\
0&-1
\end{pmatrix},\label{Mcurrent2}
\end{equation}
in terms of a matrix $\tilde{\cal M}_{m}$ related to ${\cal M}_{m}$ by a similarity transformation,
\begin{equation}
\tilde{\cal M}_{m}={\cal R}{\cal M}_{m}{\cal R}^{-1},\;\;
{\cal R}=2^{-1/2}\begin{pmatrix}
{\cal J}^{1/2}&{\cal J}^{1/2}\\
{\cal J}^{1/2}&-{\cal J}^{1/2}
\end{pmatrix}.\label{MMtilde2}
\end{equation}
Eq.\ \eqref{Mcurrent2} follows only from Eqs.\ \eqref{currentconservation} and \eqref{Jxdef} if the matrix ${\cal R}$ is Hermitian, which it is since ${\cal J}$ is Hermitian with only positive eigenvalues [see Eq.\ \eqref{jldef}].

It now follows directly from Eq.\ \eqref{currentconservation} that the matrix $U_{m}$ constructed from $\tilde{\cal M}_{m}$ by
\begin{equation}
\tilde{\cal M}_{m}=\begin{pmatrix}
a&b\\
c&d
\end{pmatrix}
\Leftrightarrow
U_{m}=\begin{pmatrix}
-d^{-1}c&d^{-1}\\
a-bd^{-1}c&bd^{-1}
\end{pmatrix}\label{YtoU}
\end{equation}
is a unitary matrix. Matrix multiplication of $\tilde{\cal M}_{m}$'s induces a nonlinear composition of $U_{m}$'s,
\begin{equation}
\tilde{\cal M}_{1}\tilde{\cal M}_{2}\Leftrightarrow U_{1}\otimes U_{2},\label{MMUU}
\end{equation}
defined by
\begin{subequations}
\label{composition}
\begin{align}
&\begin{pmatrix}
A_1&B_1\\
C_{1}&D_{1}
\end{pmatrix}
\otimes
\begin{pmatrix}
A_2&B_2\\
C_{2}&D_{2}
\end{pmatrix}
=\begin{pmatrix}
A_3&B_3\\
C_3&D_3
\end{pmatrix},\\
&A_3=A_1+B_1(1-A_2D_1)^{-1}A_2C_1,\\
&B_3=B_1(1-A_2D_1)^{-1}B_2,\\
&C_3=C_2(1-D_1A_2)^{-1}C_1,\\
&D_3=D_2+C_2(1-D_1A_2)^{-1}D_1B_2.
\end{align}
\end{subequations}

To evaluate the product \eqref{MMmrelation} of ${\cal M}_{m}$'s in a stable way, we first write it in terms of the matrices $\tilde{\cal M}_{m}$,
\begin{equation}
{\cal M}={\cal R}^{-1}\left(\prod_{m=1}^{M}\tilde{\cal M}_{m}\right){\cal R}.\label{MMmrelation2}
\end{equation}
We then transform each transfer matrix $\tilde{\cal M}_{m}$ into a unitary matrix $U_{m}$ according to Eq.\ (\ref{YtoU}) and we compose the unitary matrices according to Eq.\ \eqref{composition}. Each step in this calculation is numerically stable. 

At the end of the calculation, we may in principle transform back from the final unitary matrix $U$ to the transfer matrix ${\cal M}={\cal R}^{-1}\tilde{\cal M}{\cal R}$ by means of the inverse of relation \eqref{YtoU},
\begin{equation}
U=\begin{pmatrix}
A&B\\
C&D
\end{pmatrix}
\Leftrightarrow
\tilde{\cal M}=\begin{pmatrix}
C-DB^{-1}A&DB^{-1}\\
-B^{-1}A&B^{-1}
\end{pmatrix}.\label{UtoY}
\end{equation}
This inverse transformation is itself unstable, but we may avoid it because [as we can see by comparing Eqs.\ \eqref{YtoU} and \eqref{UtoY} with Eq.\ \eqref{ttos}] the final $U$ is \textit{identical} to the scattering matrix $S$ between leads in the infinite wave vector limit. Hence the conductance can be directly obtained from $U$ via the Landauer formula \eqref{GG0} [with the $T_{n}$'s being the eigenvalues of $BB^{\dagger}$ and $CC^{\dagger}$].

\section{Crossover from ballistic to diffusive conduction}
\label{crossover}

Away from the Dirac point (for Fermi wave vectors $k_{F}=|E-U_{0}|/\hbar v$ in the strip large compared to $1/L$) conduction through the strip is via propagating rather than evanescent modes. If the number $N_{\rm strip}=k_{F}W/\pi$ of propagating modes is $\gg 1$, the semiclassical Boltzmann equation can be used to calculate the conductance.

As the transport mean free path $l_{0}$ is reduced by adding disorder to the strip, the conduction crosses over from the ballistic to the diffusive regime. How to describe this crossover is a well-known problem in the context of radiative transfer \cite{Paa08}. An exact solution of the Boltzmann equation does not provide a closed-form expression for the crossover, but the following formula has been found to be accurate within a few percent:
\begin{equation}
\langle G\rangle=C_{d}G_{0}N_{\rm strip}\frac{l_{0}}{L+2\xi}.\label{Gxi}
\end{equation}
The coefficient $C_{d}$ depends on the dimensionality $d$: $C_{3}=4/3$, $C_{2}=\pi/2$, $C_{1}=2$. The length $\xi$ is the socalled extrapolation length of radiative transfer theory, equal to $l_{0}$ times a numerical coefficient that depends on the reflectivity of the interface at $x=0$ and $x=L$. An infinite potential step in the Dirac equation has $\xi=l_{0}$, see Ref.\ \cite{Sep08}. Substitution into Eq.\ \eqref{Gxi} then gives the formula \eqref{mfp} used in the text.


\begin{thebibliography}{99}
\bibitem{note1} Symplectic symmetry also implies that the basis of modes in the leads can be chosen such that $S$ is antisymmetric ($S=-S^{T}$). We use a different basis, so our $S$ will be not be antisymmetric. For a direct proof of the Kramers degeneracy of the transmission eigenvalues from the antisymmetry of the scattering matrix, see J. H. Bardarson, J. Phys. A \textbf{41}, 405203 (2008).
\end{thebibliography}
\end{document}